\documentclass[]{aastex631}
\shorttitle{AASTeX v6.3.1 Sample article}
\shortauthors{Ding et al.}
\graphicspath{{./}{figures/}}

\begin{document}

\title{Fundamental Parameters of a Binary System Consisting of a Red Dwarf and a Compact Star}


\correspondingauthor{Xu Ding, KaiFan Ji, ChuanJun Wang}
\email{dingxu409@163.com, jkf@ynao.ac.cn, wcj@ynao.ac.cn}

\author{Xu Ding}
\affiliation{Yunnan Observatories, Chinese Academy of Sciences (CAS), P.O. Box 110, 650216 Kunming, P. R. China}
\affiliation{Key Laboratory of the Structure and Evolution of Celestial Objects, Chinese Academy of Sciences, P. O. Box 110, 650216 Kunming, P. R. China}
\affiliation{Center for Astronomical Mega-Science, Chinese Academy of Sciences, 20A Datun Road, Chaoyang District, Beijing, 100012, P. R. China}

\author{KaiFan Ji}
\affiliation{Yunnan Observatories, Chinese Academy of Sciences (CAS), P.O. Box 110, 650216 Kunming, P. R. China}
\affiliation{Key Laboratory of the Structure and Evolution of Celestial Objects, Chinese Academy of Sciences, P. O. Box 110, 650216 Kunming, P. R. China}
\affiliation{Center for Astronomical Mega-Science, Chinese Academy of Sciences, 20A Datun Road, Chaoyang District, Beijing, 100012, P. R. China}
\affiliation{University of the Chinese Academy of Sciences, Yuquan Road 19\#, Shijingshan Block, 100049 Beijing, P.R. China}

\author{ZhiMing Song}
\affiliation{School of Information, Yunnan University of Finance and Economics, Kunming, China.}
\affiliation{Yunnan Key Laboratory of Service Computing, Kunming, China.}

\author{NianPing Liu}
\affiliation{Yunnan Observatories, Chinese Academy of Sciences (CAS), P.O. Box 110, 650216 Kunming, P. R. China}
\affiliation{Key Laboratory of the Structure and Evolution of Celestial Objects, Chinese Academy of Sciences, P. O. Box 110, 650216 Kunming, P. R. China}
\affiliation{Center for Astronomical Mega-Science, Chinese Academy of Sciences, 20A Datun Road, Chaoyang District, Beijing, 100012, P. R. China}

\author{JianPing Xiong}
\affiliation{Yunnan Observatories, Chinese Academy of Sciences (CAS), P.O. Box 110, 650216 Kunming, P. R. China}
\affiliation{Key Laboratory of the Structure and Evolution of Celestial Objects, Chinese Academy of Sciences, P. O. Box 110, 650216 Kunming, P. R. China}
\affiliation{Center for Astronomical Mega-Science, Chinese Academy of Sciences, 20A Datun Road, Chaoyang District, Beijing, 100012, P. R. China}

\author{Qiyuan Cheng}
\affiliation{Yunnan Observatories, Chinese Academy of Sciences (CAS), P.O. Box 110, 650216 Kunming, P. R. China}
\affiliation{Key Laboratory of the Structure and Evolution of Celestial Objects, Chinese Academy of Sciences, P. O. Box 110, 650216 Kunming, P. R. China}
\affiliation{Center for Astronomical Mega-Science, Chinese Academy of Sciences, 20A Datun Road, Chaoyang District, Beijing, 100012, P. R. China}
\affiliation{University of the Chinese Academy of Sciences, Yuquan Road 19\#, Shijingshan Block, 100049 Beijing, P.R. China}

\author{ChuanJun Wang}
\affiliation{Yunnan Observatories, Chinese Academy of Sciences (CAS), P.O. Box 110, 650216 Kunming, P. R. China}
\affiliation{Key Laboratory of the Structure and Evolution of Celestial Objects, Chinese Academy of Sciences, P. O. Box 110, 650216 Kunming, P. R. China}
\affiliation{Center for Astronomical Mega-Science, Chinese Academy of Sciences, 20A Datun Road, Chaoyang District, Beijing, 100012, P. R. China}
\affiliation{University of the Chinese Academy of Sciences, Yuquan Road 19\#, Shijingshan Block, 100049 Beijing, P.R. China}

\author{JinLiang Wang}
\affiliation{Yunnan Observatories, Chinese Academy of Sciences (CAS), P.O. Box 110, 650216 Kunming, P. R. China}
\affiliation{Key Laboratory of the Structure and Evolution of Celestial Objects, Chinese Academy of Sciences, P. O. Box 110, 650216 Kunming, P. R. China}
\affiliation{Center for Astronomical Mega-Science, Chinese Academy of Sciences, 20A Datun Road, Chaoyang District, Beijing, 100012, P. R. China}
\affiliation{University of the Chinese Academy of Sciences, Yuquan Road 19\#, Shijingshan Block, 100049 Beijing, P.R. China}

\author{DeQing Wang}
\affiliation{Yunnan Observatories, Chinese Academy of Sciences (CAS), P.O. Box 110, 650216 Kunming, P. R. China}
\affiliation{Key Laboratory of the Structure and Evolution of Celestial Objects, Chinese Academy of Sciences, P. O. Box 110, 650216 Kunming, P. R. China}
\affiliation{Center for Astronomical Mega-Science, Chinese Academy of Sciences, 20A Datun Road, Chaoyang District, Beijing, 100012, P. R. China}

\author{ShouSheng He}
\affiliation{Yunnan Observatories, Chinese Academy of Sciences (CAS), P.O. Box 110, 650216 Kunming, P. R. China}
\affiliation{Key Laboratory of the Structure and Evolution of Celestial Objects, Chinese Academy of Sciences, P. O. Box 110, 650216 Kunming, P. R. China}
\affiliation{Center for Astronomical Mega-Science, Chinese Academy of Sciences, 20A Datun Road, Chaoyang District, Beijing, 100012, P. R. China}



\begin{abstract}
TIC 157365951 has been classified as a $\delta$ Scuti type by the International Variable Star Index (VSX). Through the spectra from Large Sky Area Multi-Object Fiber Spectroscopic Telescope (LAMOST) and its light curve, we further discovered that it is a binary system. This binary system comprises a red dwarf star and a compact star. Through the spectral energy distribution (SED) fitting, we determined the mass of the red dwarf star as $M_1 = 0.31 \pm 0.01 M_{\odot}$ and its radius as $R_1 = 0.414 \pm 0.004 R_{\odot}$. By fitting the double-peaked H${\rm \alpha}$ emission, we derived the mass ratio of $q = 1.76 \pm 0.04 $, indicating a compact star mass of $M_2 = 0.54 \pm 0.01 M_{\odot}$. Using Phoebe to model the light curve and radial velocity curve for the detached binary system, we obtained a red dwarf star mass of $M_1 = 0.29 \pm 0.02 M_{\odot}$, a radius of $R_1 = 0.39 \pm 0.04 R_{\odot}$, and a Roche-lobe filling factor of $f = 0.995\pm0.129$, which is close to the $f=1$ expected for a semi-detached system. The Phoebe model gives a compact star mass $M_2 = 0.53 \pm 0.05 M_{\odot}$.
Constraining the system to be semidetached gives $M_1 = 0.34 \pm 0.02 M_{\odot}$, $R_1 = 0.41 \pm 0.01 R_{\odot}$, and $M_2 = 0.62 \pm 0.03 M_{\odot}$. The consistency of the models is encouraging. The value of the Roche-lobe filling factor suggests that there might be ongoing mass transfer. The compact star mass is as massive as a typical white dwarf. 
\end{abstract}

\keywords{White dwarf stars (1799) --- Compact objects (288) --- Ellipsoidal variable stars (455)}


\section{Introduction} 
Most stars conclude their lifecycles as compact objects, which include white dwarfs (WDs), neutron stars (NSs), and black holes (BHs). The evolutionary pathways leading to these endpoints are influenced by a variety of factors, such as the initial mass of the star, its metallicity, and its rotation rate, etc. In recent years, there has been a surge of interest in locating non-interacting binaries housing a compact object \citep{Breivik+et+al+2017}. While these investigations often concentrate on pinpointing black hole binaries \citep{Chakrabarti+et+al+2023, El-Badry+et+al+2023, Tanikawa+et+al+2023}, the identical set of astrometric \citep{Andrews+et+al+2019}, spectroscopic \citep{Jayasinghe+et+al+2023}, and photometric tools \citep{ Rowan+et+al+2021,Green+et+al+2023} have been employed to identify potential neutron stars \citep{Lin+et+al+2023, Zheng+et+al+2023} and white dwarfs \citep{Rowan+et+al+2024} as well .

Stellar-mass black holes and neutron stars are typically identified through X-ray emissions, a result of substantial gas accretion \citep{Kreidberg+et+al+2012, Liu+et+al+2013}. Various methods are utilized to detect individual compact objects or those in binary systems. Radial velocity (RV) monitoring stands out as particularly effective for uncovering massive yet unseen compact companions within binary systems, especially when coupled with extensive spectroscopic surveys. Systems identified through RV monitoring typically exhibit a quiescent phase with exceedingly faint (potentially undetectable) X-ray emissions \citep{Yang+et+al+2021}.

Here, we illustrate that the subject named TIC 157365951 is a binary system through RV observations, featuring a red dwarf and a compact object. This compact star may be a white dwarf. This paper is organized as follows: Section 2 outlines our observations, encompassing photometric and spectroscopic details. Section 3 delves into the characterization of the binary. Finally, the concluding section offers discussion and conclusions drawn from the study's findings.

\section{Observations}
\subsection{Photometry}
The TESS instrument comprises four wide-field CCD cameras, each boasting a 24° × 24° field of view and a 10.5 cm entrance pupil diameter. 
The cameras, arranged in a 24° × 96° rectangle, operate within a passband range of 600 to 1000 nm \citep{Ricker+et+al+2015}. TESS, operational since April 2018, aims to deliver precise photometry for over 200,000 stars, allocating a minimum observation time of 27 days per star with a cadence ranging with 2 or 30 minutes. TIC 157365951 underwent four observations by the TESS telescope across sectors 23, 24, 50, and 51. We conducted a comprehensive survey of all available light curves with with 2-minute cadence for this star from TESS, sourced from the Mikulski Archive for Space Telescopes (MAST) server. The data underwent processing via the Science Processing Operations Center (SPOC) pipeline \citep{Jenkins+et+al+2016}. Two flux types were accessible: Simple Aperture Photometry (SAP) and Pre-search Data Conditioning SAP (PDCSAP). We chose to utilize the PDCSAP flux for our analysis due to its generally higher cleanliness compared to the SAP flux \citep{Jenkins+et+al+2010}. Figure 1 illustrates the light curve for each sector, revealing distinct flare events across all sectors. Notably, two significant outbursts were observed in sectors 23 and 51, respectively.

\begin{figure}[!ht]
\begin{center}

\begin{minipage}{12cm}
	\includegraphics[width=10cm]{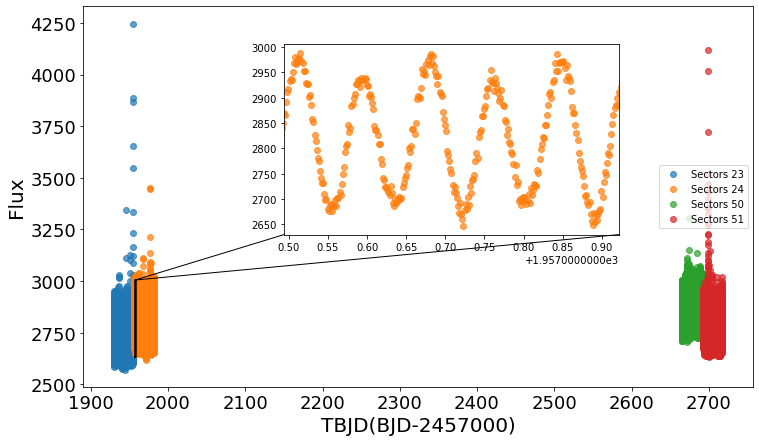}
\end{minipage}

\caption{Each sector’s observed light curves revealed notable flare events.}\label{Fig1}
\end{center}
\end{figure}

\subsection{Spectroscopy}
The Large Sky Area Multi-Object Fiber Spectroscopic Telescope (LAMOST) is a reflective Schmidt telescope boasting a 4-meter aperture and 5-degree field of view. Its focal plane, spanning a diameter of 1.75 meters, is outfitted with 4000 fibers, distributed among 16 spectrographs, with each capable of accommodating up to 250 fibers. This setup allows LAMOST to simultaneously capture the spectra of up to 4000 targets, achieving a resolution of R $\sim$ 1800. After a 1.5-hour exposure, it can detect objects down to around r $\sim$ 18 magnitude. In October 2018, LAMOST initiated its 5-year medium-resolution spectroscopic (MRS) survey, characterized by a spectral resolution of R $\sim$ 7500 and a limiting magnitude of G $<$ 15 mag. The survey incorporates a time-domain spectroscopic subproject, envisioned to encompass approximately $\sim$ 60 exposures spanning five years. Its objective is to observe more than 100,000 stars \citep{Liu+et+al+2020}, offering a distinctive chance to gather a significant sample of binary stars with resolved orbits. TIC 157365951 was observed with the LAMOST telescope, and we obtained medium-resolution spectral data from the DR10 dataset.

\section{Binary Characterization} 

\subsection{Period determination}
In order to determine the orbital period of TIC 157365951, we employed the Lomb-Scargle  \citep{Lomb+et+al+1976, Scargle+et+al+1982} Periodograms to obtain the power spectrum of the target which were implemented in the Astropy package \citep{Astropy+et+al+2022}. To enhance the precision of the period estimation, we employ a lengthy time series spanning from all the sectors to compute the Lomb–Scargle power spectrum, thereby improving its accuracy. We assess the period uncertainty utilizing a bootstrap technique \citep{Efron+et+al+1979}, executing it across 10,000 iterations, wherein random points are selectively excluded from each iteration. From the Lomb–Scargle periodogram analysis, the estimated period is determined with an error margin of $P_{\rm orb} = 0.16756(1)$ days. It’s worth noting that for the ellipsoidal variation, the actual orbital period is twice the dominant period identified on the Lomb–Scargle power spectrum. We define phase zero as the lower of the two minima in the light curve, for which the parabolic fit yields $T_0 = 2457000.0 ~ BJD + 1929.57121(5)$. The folded light curve with phase-flux based on the $P_{\rm orb}$ and $T_0$ is shown in the Figure 2.

\begin{figure}[!ht]
\begin{center}
\begin{minipage}{10cm}
	\includegraphics[width=10cm]{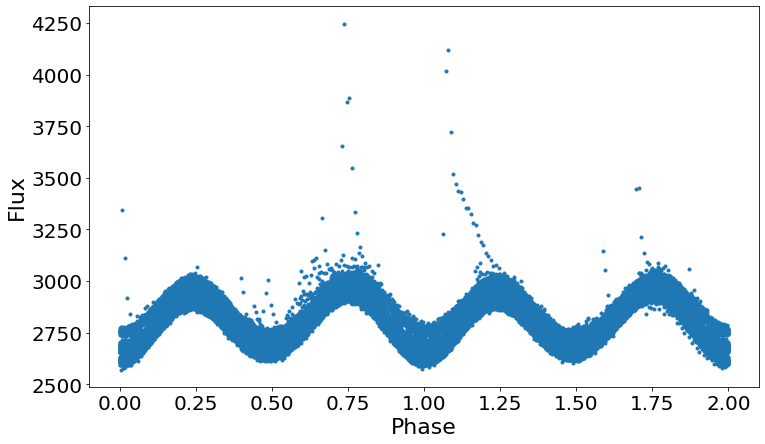}
\end{minipage}
\caption{The folded light curve with phase-flux is shown.}\label{Fig2}
\end{center}
\end{figure}

\subsection{H\texorpdfstring{$\alpha$}{alpha} emission}
 The spectral type of TIC 157365951 is M3.5V according to Simbad. The medium-resolution spectral data from LAMOST show variable H$\alpha$ emission in Figure 3. The orbital phase is characterized by the point where the maximum radial velocity (RV) of the M-type star is observed at phase = 0.75, while the M-type star is positioned behind the companion star at phase = 0. The next section discusses that the zero phase of radial velocity corresponds to an M-dwarf star being behind the companion star. H$\alpha$ emission may arise from a combination of chromospheric activity and/or mass transfer. For instance, double-peaked emission is frequently observed in accreting compact object binaries \citep{Swihart+et+al+2022}. The double-peaked emission line spectrum primarily occurs near phases 0.25 and 0.75. The H$\alpha$ emission seems to correlate with the movement of the M-dwarf, aligning with chromospheric emission observed in analogous binaries featuring compact companions \citep{Lin+et+al+2023, Zheng+et+al+2023}. The radial velocity of the narrow component (discussed further below) varies in phase with the expected motion of the compact object. However, its physical origin is not fully explained, as material accreting onto a compact object would generally be expected to exhibit higher infall or orbital velocities.

 \subsection{The spectroscopic orbit of the M-dwarf}
The radial velocity curves were derived by the code developed by \citet{Zhang+et+al+2021} based on the cross-correlation function method. To circumvent issues related to the H$\alpha$ region and its associated activity, we exclusively utilized spectra from the blue arm for radial velocity determination. The left panel of Figure 4 shows the blue-arm velocities of M-type dwarf as a function of phase. The RVs were fitted using a circular orbit model as per the equation
\begin{equation}
    V(\phi) = -K_1\sin[\omega(\phi + \Delta \phi)] + \gamma
\end{equation}

where $K_1$ represents the semi-amplitude of the RVs of the visible star, \edit1{$\omega = 2\pi/P_{\rm orb}$}, $\gamma$ denotes the systemic velocity of TIC 157365951 and $\Delta \phi$ accounts for the potential zero-point shift phase attributed to the limited accuracy in period and $T_0$ determination when folding the RVs. The derived fitting results are $K_1 = 168 \pm 3 $ km/s,
$\gamma = -17 \pm 2$ km/s, and \edit1{$\Delta \phi = 0.039 \pm 0.002$}.

The mass function of a binary system is defined as
\begin{equation}
    f(M_2) = \frac{M_2^3\sin^3i}{(M_1+M_2)^2} = \frac{K_1^3P_{{\rm orb}}}{2\pi G}
\end{equation}

Where $M_1$ and $M_2$ are the masses of the visible star and the compact star respectively, $K_1$ is the semi-amplitude of the radial velocities (RVs) of the visible star, $P_{\rm orb}$ is the orbital period, and $i$ is the orbital inclination of the binary system to the observer. This function provides the minimum possible mass of the compact star. Given $P_{\rm orb} = 0.16756(1)$ days and $K_{\rm p} = 168 \pm 3 $ km/s, we obtain the mass function of the compact star, $f(M_2) = 0.083 \pm 0.002 M_{\odot}$.

The occurrence of the double-peaked emission line spectrum (H$\alpha$) is predominantly observed at specific phases. The observed bimodal peaks appear near the quadrants of maximum velocity difference between the two stars, from phase 0.115 to 0.309 on one side of the orbit, and from phase 0.633 to 0.830 on the other side, as indicated in red in Figure 3. In Figure 3, the horizontal axis represents wavelengths in vacuum, and the vacuum wavelength of H$\alpha$ is 6564.61 $\text{\AA}$. During each phase, the approach of fitting (H$\alpha$) profiles was employed by utilizing the sums of two Gaussians \citep{Kjurkchieva+et+al+2003}. The radial velocity curves are shown in the right panel of Figure 4, with red and blue dots representing them respectively. The green points represent radial velocities derived from the blue arm. The blue dots and green dots overlap well. We fitted the red dots using Equation 1. The derived fitting results are $K_{\rm s} = -95 \pm 2 $ km/s,
$\gamma = -17 \pm 2$ km/s, and $\Delta \phi = 0.045 \pm 0.007$. 
Under the assumption that the velocities of the extra component of the H$\alpha$ emission line (red dots) accurately reflect the velocity of the center of mass of the compact object, the system mass ratio is $q = M_2/M_1 = K_{\rm p}/K_{\rm s} = 1.76 \pm 0.04$.

\begin{figure}[!ht]
\begin{center}
\begin{minipage}{8cm}
	\includegraphics[width=8cm]{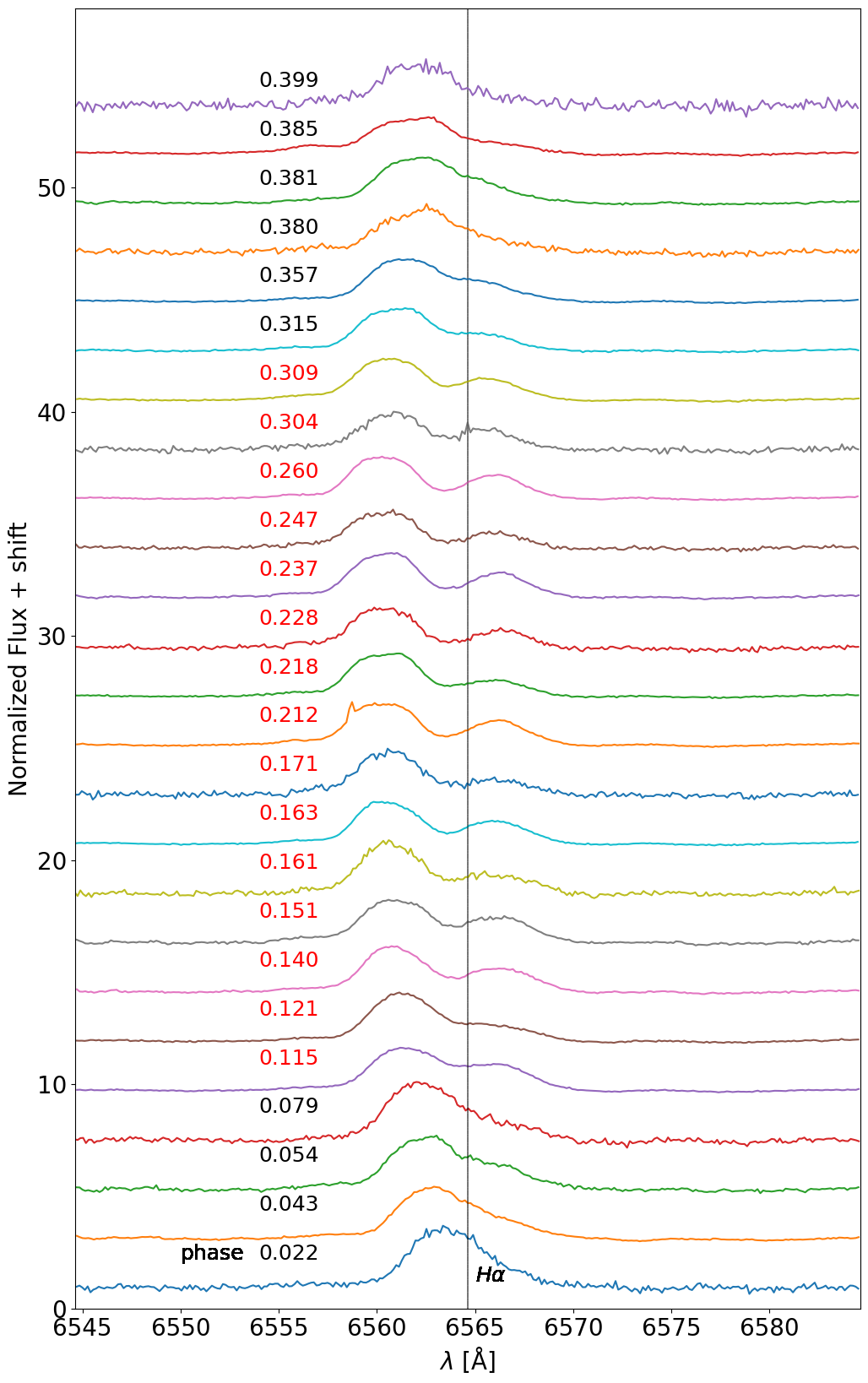}
\end{minipage}
\begin{minipage}{8cm}
	\includegraphics[width=8cm]{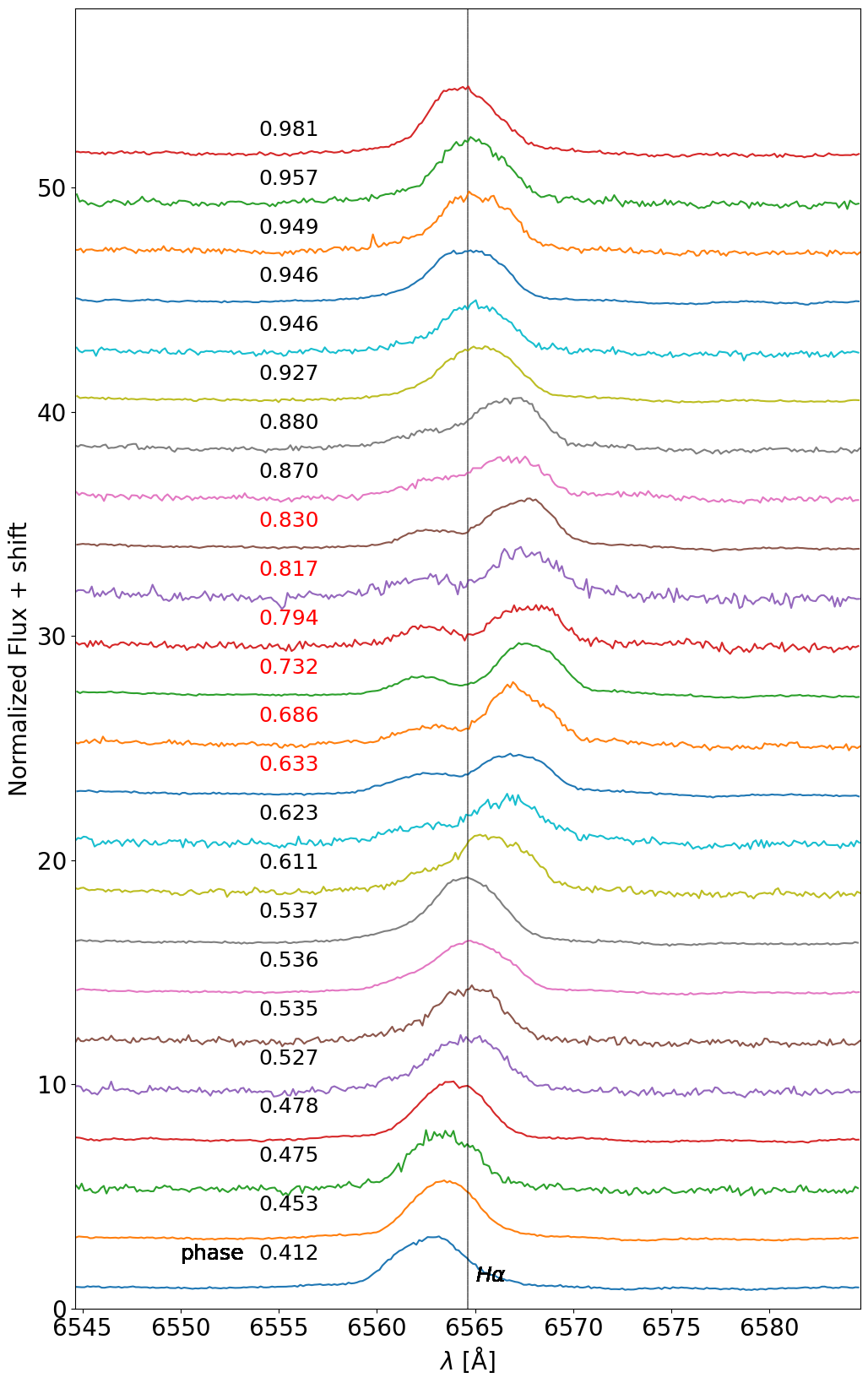}
\end{minipage}
\caption{The  H$\alpha$ spectra of TIC 157365951 are shown alongside a dashed line marking the position of the  H$\alpha$ line.}\label{Fig3}
\end{center}
\end{figure}

\begin{figure}[!ht]
\begin{center}
\begin{minipage}{8cm}
	\includegraphics[width=8cm]{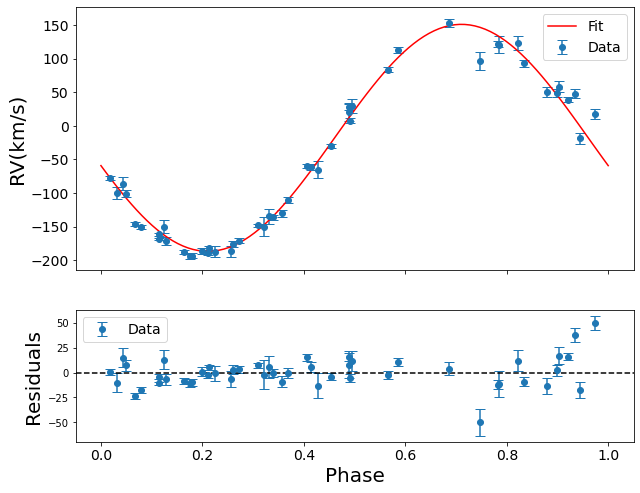}
\end{minipage}
\begin{minipage}{8cm}
	\includegraphics[width=8cm]{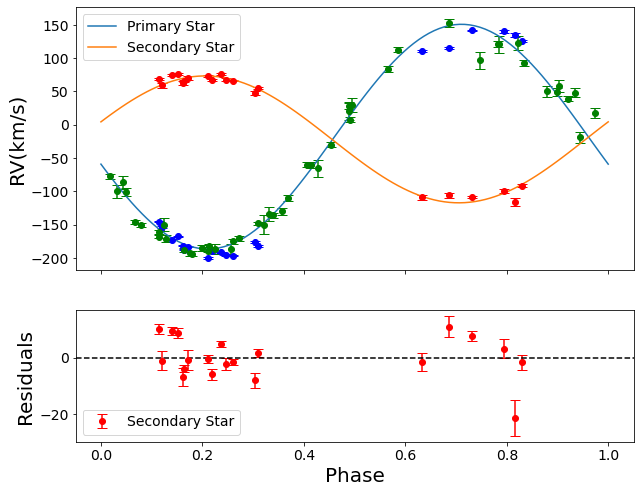}
\end{minipage}
\caption{Left: the radial velocities of M-type dwarfs were only obtained from the blue arm. Right: the red and blue points represent the radial velocity curves calculated using H$\alpha$. The green points mirror those in the left panel.}\label{Fig4}
\end{center}
\end{figure}

\subsection{Spectral energy distributions}

Independent estimates of the parameters of the visible star can be provided by the spectral energy distribution (SED). To perform Spectral Energy Distribution (SED) fitting, we employ the SPEEDYFIT package \footnote{https://github.com/vosjo/speedyfit}. This tool leverages \citet{Kurucz+et+al+1979} models and a Markov Chain Monte Carlo (MCMC) method to identify the global minimum and calculate the uncertainties of the fitting parameters. The SPEEDYFIT code internally calculates a solid angle based on flux and effective temperature. Given the distance, determined from the Gaia parallax, this solid angle is then converted into a physical radius. The fit also provides a surface gravity ($g$), which, combined with the radius ($R$), allows for the derivation of the mass ($M = gR^2/G$), where $G$ is the gravitational constant. We impose limitations on the solution by incorporating the Gaia DR3 \citep{Gaia+et+al+2023} parallax and surface gravity, where $\varpi = 27.10 \pm 0.01$ mas and $\log g = 4.7$. We utilize published photometry whenever feasible. SPEEDYFIT excludes ZTF and TESS filters from its compatibility list, thus they are not utilized in our analysis. Errors in the constraints are incorporated as priors in the Bayesian fitting process, consequently propagating them to the final results for examination. A more comprehensive explanation of the fitting procedure is provided in the work by \citet{Vos+et+al+2017}. The posterior distribution and fitted results are depicted in Figure 5 and Figure 6, with a detailed listing of parameters provided in Table 1. The effective temperature reported by Gaia DR3 \citep{Gaia+et+al+2023} is 3186K. Additionally, the effective temperature derived from the LAMOST spectral data is 3282K by LAMOST Stellar Parameter Pipeline (LASP) \citep{Luo+et+al+2015}. The temperature obtained from SED fitting deviates slightly from these temperatures. The reason for the temperature falling at the edge of the distribution is due to the use of \citet{Kurucz+et+al+1979} models using SPEEDYFIT package, which can only be set with a lower limit of 3500K. The given temperature and uncertainty are rather meaningless, so Table 1 does not list it. The mass of the visible star, obtained through SED fitting, is $M_1 = 0.31 \pm 0.01$$M_{\odot}$, and the mass ratio $q = 1.76 \pm 0.04 $, determined through fitting Ha, yields a companion  star mass of $M_2 = M_1 \times q = 0.54 \pm 0.01$$M_{\odot}$. The solution is constrained using the Gaia DR3 parallax ($\varpi = 27.10 \pm 0.01$ mas), surface gravity ($\log g = 4.7$), atmosphere model, and published photometry where available. The deviations in observational measurements and the temperature limitations of atmosphere model both impact the calculations of mass.

\begin{figure}[!ht]
\begin{center}
\begin{minipage}{13cm}
	\includegraphics[width=13cm]{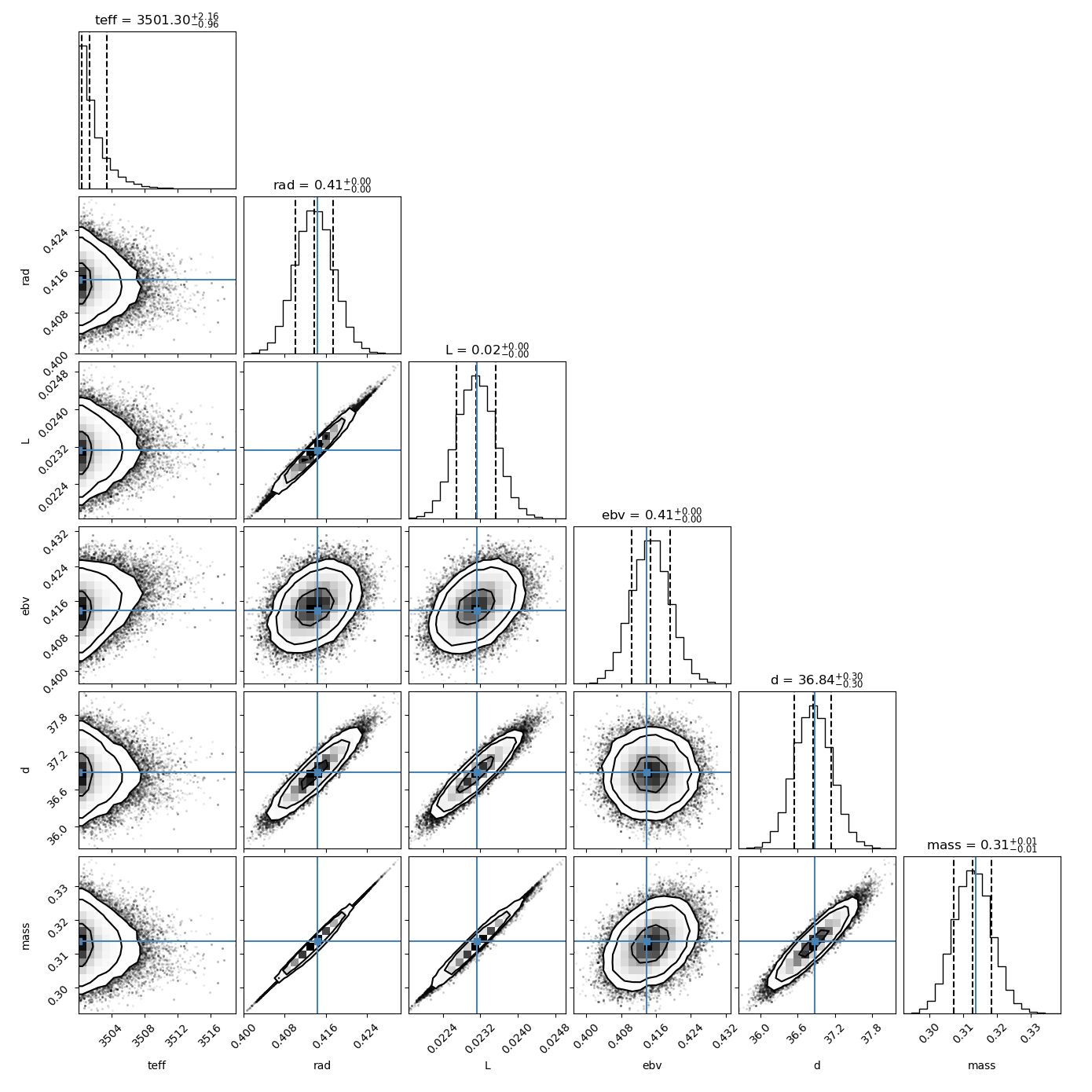}
\end{minipage}
\caption{The distributions of parameters from SED fitting.}\label{Fig5}
\end{center}
\end{figure}

\begin{figure}[!ht]
\begin{center}
\begin{minipage}{8cm}
	\includegraphics[width=8cm]{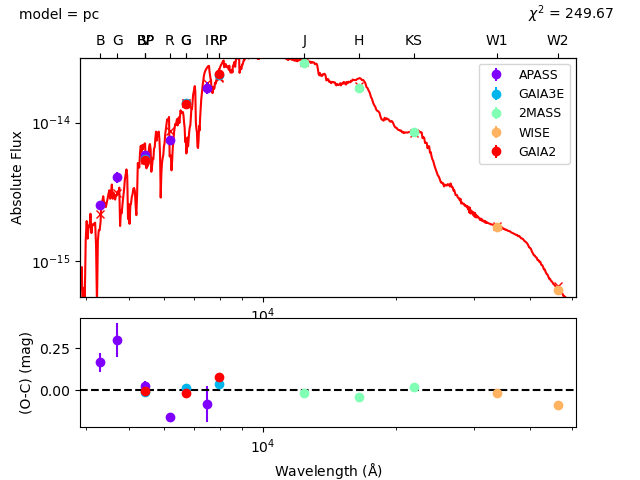}
\end{minipage}
\caption{SED fitting using SPEEDYFIT. The upper panels display the observations and spectral energy distributions of the visible star. The residuals of the fit are presented in the lower panel.}\label{Fig6}
\end{center}
\end{figure}

\subsection{Light curve Modeling}
When studying M-dwarf, we first analyzed the M-dwarf without constraining its radius to match the Roche critical lobe, and then added that constraint, \edit1{to assess whether mass transfer is ongoing.}

The analysis begins with a detached binary system of M-dwarf that has not yet filled its Roche lobe. Figure 1 displays the light curve, exhibiting the O’Connell effect  \citep{Connell+et+al+1951, Milone+et+al+1968}, characterized by the uneven height between the maximum at 0.25 phase and the maximum at 0.75 phase. Given the pronounced magnetic activity of the target, which significantly influences the shape of the light curve, we focused solely on folding the first three periods of the time-flux data from sectors 23 into phase-magnitude data. In this study, we introduce the incorporation of cool spots to elucidate this phenomenon exhibiting the O’Connell effect. To ascertain the fundamental parameters of a binary system, a crucial initial step involves presuming the effective temperature of the visible star. Based on the LAMOST spectral data, we derived a temperature of 3282K for the visible star. In the detached binary configuration, we derive the posterior parameter distribution utilizing Phoebe 2.3 \citep{prsa+et+al+2005, prsa+et+al+2011, prisa+et+al+2016, Horvat+et+al+2018, Jones+et+al+2020, Conroy+et+al+2020} in conjunction with Markov Chain Monte Carlo (MCMC) techniques. If one component’s effective temperature is below 6600K, Phoebe suggests setting the gravity-darkening coefficient ($g$) to 0.32 and the bolometric albedo (${\it A}$) to 0.6. The parameters for the visible star have been adjusted to $g_1 = 0.32$ and $A_1=0.6$. We categorize the companion star within the system as an obscure companion, specifying its dimensions as minuscule ($R_2 = 3 \times 10^{-6} R_{\odot}$) and its temperature as notably low ($T_2 = 300 K$). We set the eclipse model using the eclipse method as ‘only horizon’, following the approach used by \citet{Jayasinghe+et+al+2021}. The TESS:T passband is employed to cover a wide spectral range spanning from 600 to 1000 nm \citep{Ricker+et+al+2015}. The orbital period is fixed at 0.16756 days, while other parameters are kept adjustable. These include the orbital inclination ($incl$), mass ratio ($q$), visible star radius ($R_1$), cool spot characteristics, and semi-major axis (sma). The defining parameters of a cool spot typically consist of its colatitude ($colat$), longitude ($long$), angular radius ($radius$), and relative temperature compared to the intrinsic local value ($relteff$).

Emcee \citep{Foreman+et+al+2019} is a software tool that employs a pure-Python implementation of the Markov chain Monte Carlo (MCMC) Ensemble sampler and is licensed under the MIT license. It is utilized to adjust the adjustable parameters (${\it incl, \it q, \it R_1, \it colat, \it long, \it radius, \it relteff, \it sma}$) mentioned in the preceding paragraph. In our modeling study, 50 walkers were employed. The autocorrelation time of our MCMC chain falls within the range of 20 to 125 steps. To ensure the convergence of the MCMC chain, it is recommended that the chain length exceeds 10-20 times the integrated autocorrelation time \citep{Conroy+et+al+2020, Li+et+al+2021}. To guarantee convergence, we set the chain length to be over 40 times the integrated autocorrelation time, resulting in a total of 5000 steps. Figures 7 and 8 display the solution outcomes. In Figure 7, the posterior distribution of the solution parameters is depicted. Figure 8 presents the results of fitting both the light curve and the radial velocity curve, where the fitted curve is generated by Phoebe utilizing the derived parameters.  

\begin{figure}[!ht]
\begin{center}
\begin{minipage}{12cm}
	\includegraphics[width=12cm]{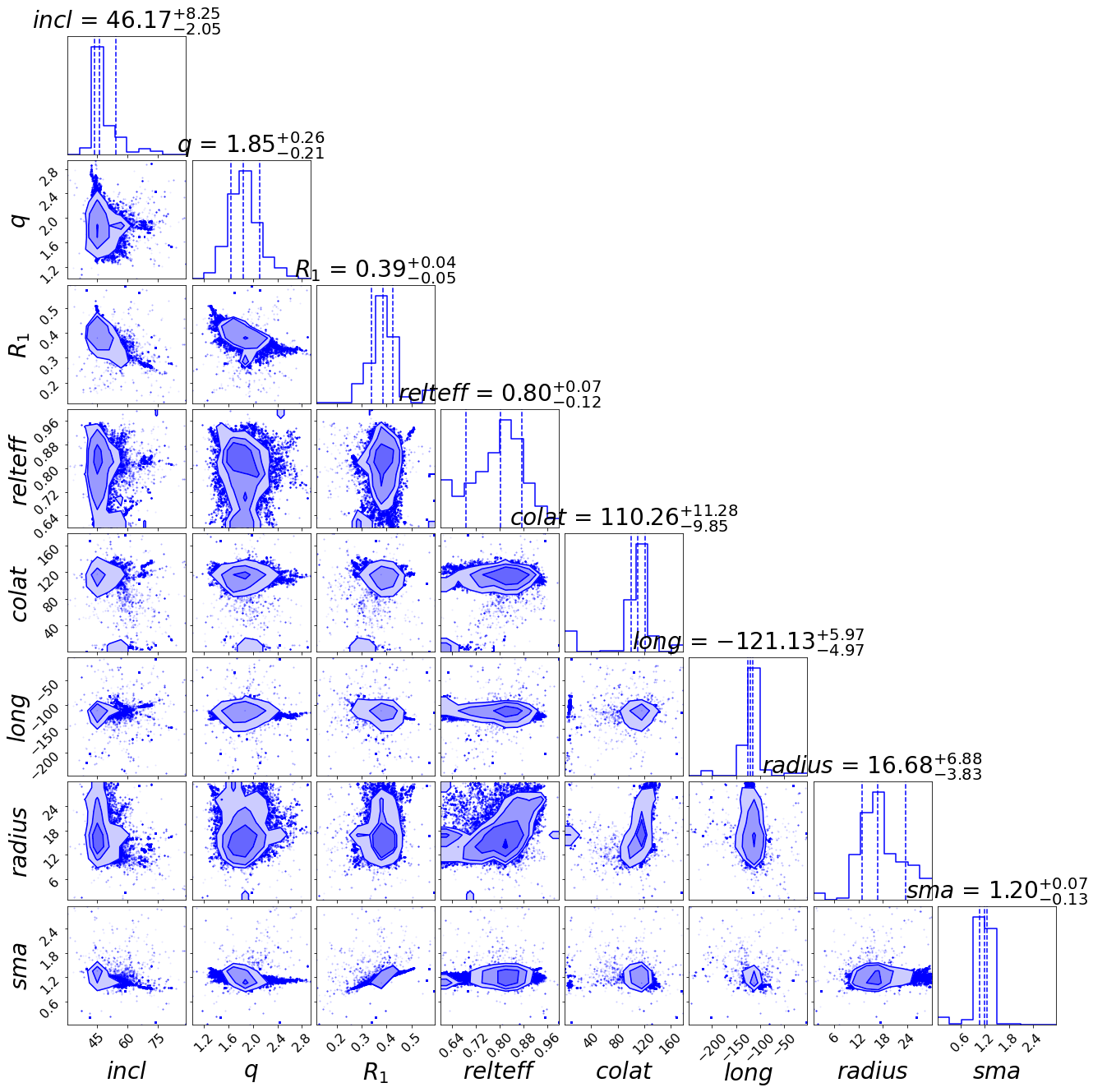}
\end{minipage}
\caption{Parameter distributions derived from the fitting results of light curve (lc) and radial velocity (RV) curve using Phoebe with a detached binary system.}\label{Fig8}
\end{center}
\end{figure}

\begin{figure}[!ht]
\begin{center}
\begin{minipage}{14cm}
	\includegraphics[width=14cm]{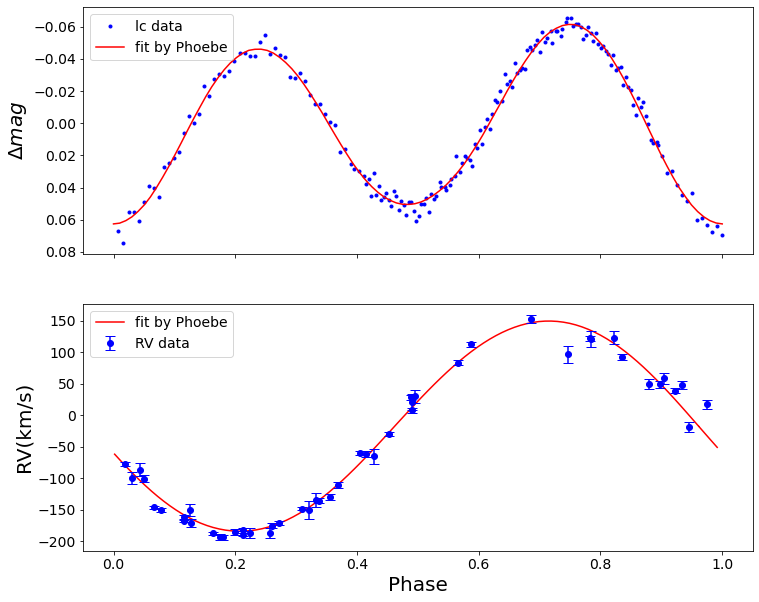}
\end{minipage}
\caption{The fitting results of light curves and radial velocity (RV) curve using Phoebe with a detached binary system.}\label{Fig9}
\end{center}
\end{figure}

We can also ascertain the Roche-lobe filling factor ($f = R_1/R_{\rm RL}$). The Roche lobe radius ($R_{\rm RL}$) is estimated using the formulation provided by \citet{Eggleton+et+al+1983}.
\begin{equation}
 \frac{R_{\rm RL}}{a} = \frac{0.49q^{-2/3}}{0.6q^{-2/3}+\ln(1+q^{-1/3})}
\end{equation}
where $a$ denotes the semimajor axis and $q$ represents the mass ratio. The Roche-lobe filling factor ($f$) is calculated as $0.995\pm0.129$. In a detached binary model, such a high Roche-lobe filling factor suggests that the M-dwarf might be filling its Roche lobe.

Guided by this result, we now analyze a semi-detached  binary system in which the M-dwarf component has already filled its Roche lobe. We fix $g_1$, $A_1$, $T_1$, and $Period$ to values consistent with those of a detached binary system. In Figure 9, we present the posterior distribution of the solution parameters. Figure 10 shows the results of fitting both the light curve and the radial velocity curve, with the fitted curve produced by Phoebe using the derived parameters.

Table 1 lists the absolute parameters generated by Phoebe based on these derived values. The results in Table 1 show that the values of $M_1$ and $R_1$ obtained from the SED solution are very close to those obtained from the modeling of both detached and semi-detached binary systems. The mass ratio obtained through Phoebe calculations closely matches that obtained through H$\alpha$ fitting. The compact star mass ($M_2$) obtained through Phoebe calculations closely matches the values obtained from SED fitting and H$\alpha$ fitting. The similarity between the parameters obtained from detached binary modeling and those from semi-detached binary modeling can be attributed to the fact that, in the detached binary model, the Roche-lobe filling factor ($f$) of the M-dwarf star reaches 0.995, which is close to 1 for a semi-detached binary system. The mass parameter obtained for $M_1$ is very similar across these methods, whereas the small discrepancy in $M_2$ might be associated with the precision of radial velocity measurements from fewer H$\alpha$ fitting points for the compact star.

\begin{figure}[!ht]
\begin{center}
\begin{minipage}{12cm}
	\includegraphics[width=12cm]{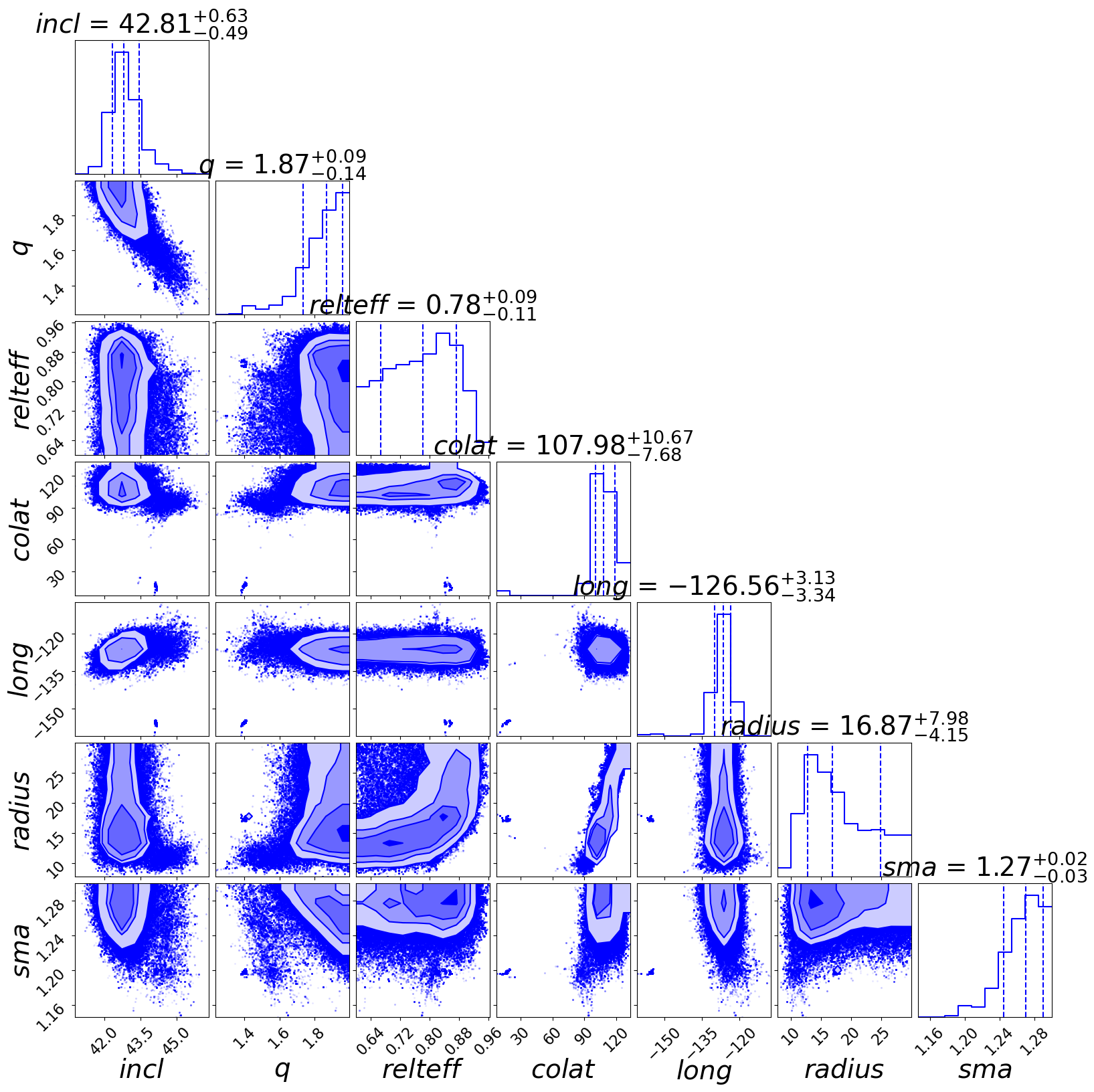}
\end{minipage}
\caption{Parameter distributions derived from the fitting results of light curve (lc) and radial velocity (RV) curve using Phoebe with a semi-detached binary system.}\label{Fig18}
\end{center}
\end{figure}

\begin{figure}[!ht]
\begin{center}
\begin{minipage}{14cm}
	\includegraphics[width=14cm]{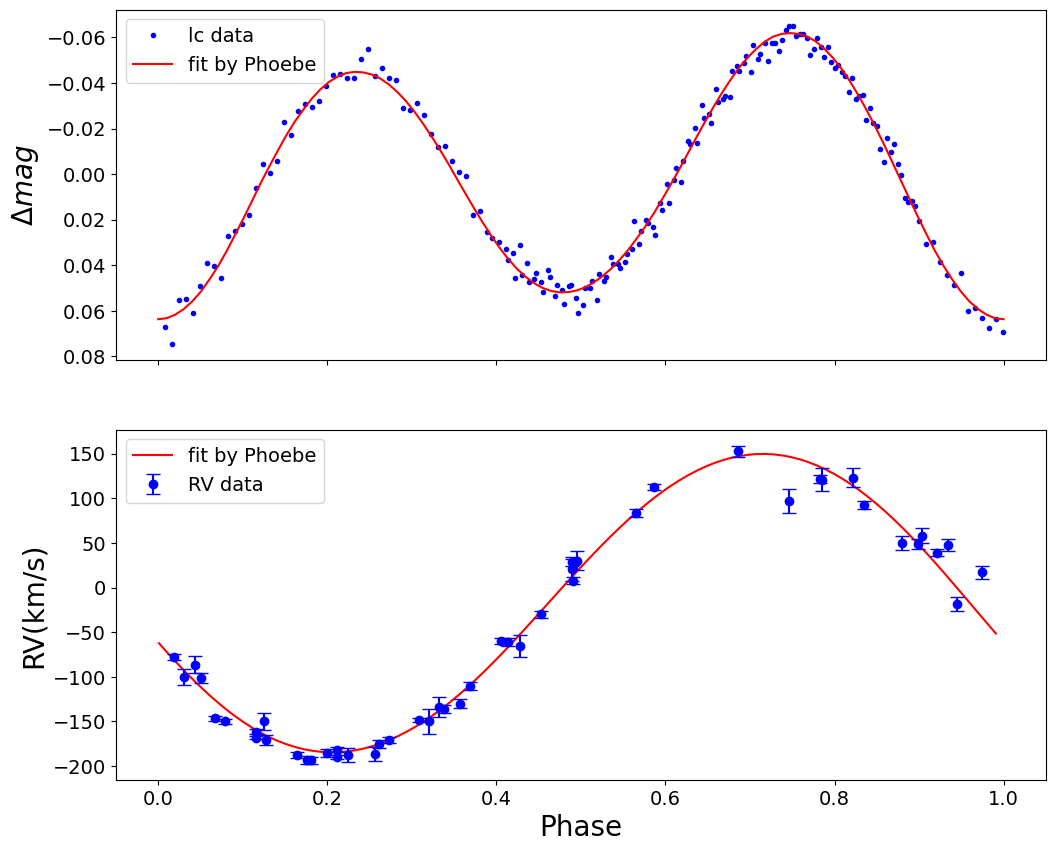}
\end{minipage}
\caption{The fitting results of light curves and radial velocity (RV) curve using Phoebe with a semi-detached binary system.}\label{Fig19}
\end{center}
\end{figure}

\clearpage
{
\tiny
\begin{center}
\begin{longtable}{ccccccc}
\caption{Fundamental parameters of TIC 157365951}\label{Table 11}\\
\hline\hline 
\textbf{Basic information(SIMBAD)} &\hspace{5cm} &     \\
RA (ICRS) &\hspace{5cm} & $15 ~ 53 ~ 04.84 $              \\
DEC (ICRS) &\hspace{5cm} & $+44 ~ 57 ~ 44.56$              \\
Spectral type &\hspace{5cm} & 	M3.5V              \\
$B~V~R$(magnitude) &\hspace{5cm} & 	$15.97 ~ 14.50 ~ 14.21$              \\
Parallaxes (pc)   &\hspace{5cm} &  $36.89 \pm 0.02$    \\
\hline
\textbf{Spectral energy distribution} &\hspace{5cm} &     \\
$R_1$ ($R_{\odot}$) &\hspace{5cm} & $0.414 \pm 0.004$              \\
$L_1$ ($L_{\odot}$) &\hspace{5cm} & $0.0231 \pm 0.0004$              \\
$E (B-V)$ (mag) &\hspace{5cm} & $0.416 \pm 0.004$              \\
$d$ (pc) &\hspace{5cm} & $36.84 \pm 0.30$              \\
$M_1$ ($M_{\odot}$) &\hspace{5cm} & $0.31 \pm 0.01$              \\
\hline
\textbf{Parameters ($H\alpha$ and SED)}   &\hspace{5cm} &     \\ 
$q(\frac{M_2}{M_1})$                 &\hspace{5cm}           &$1.76 \pm 0.04$               \\
$M_2$ ($M_{\odot}$) &$q \times M_1(SED)$           & $0.54 \pm 0.01$                         \\
\hline
\textbf{Light curve Modeling} &\hspace{5cm} &  \textbf{detached model}   \\
$g_1$           &\hspace{5cm}           &0.32(fixed)                          \\
$A_1$       &\hspace{5cm}               &0.6(fixed)                         \\
$T_1(K)$       &\hspace{5cm}                &3282(fixed)                        \\
$Period(day)$       &\hspace{5cm}           &0.16756(fixed)                          \\

$incl$($^{\circ}$)     &\hspace{5cm}           &$46.17_{-2.05}^{+8.25}$                    \\
$q(\frac{M_2}{M_1})$                 &\hspace{5cm}           &$1.85_{-0.21}^{+0.26}$               \\
$R_1$ ($R_{\odot}$) &\hspace{5cm} & $0.39 \pm 0.04$              \\

$relteff$           &\hspace{5cm}           & $0.80_{-0.12}^{+0.07}$                            \\
$colat$($^{\circ}$)     &\hspace{5cm}           & $110.26_{-9.85}^{+11.28}$                             \\
$long$($^{\circ}$)      &\hspace{5cm}           & $-121.13_{-4.97}^{+5.97}$                              \\
$radius$($^{\circ}$)            &\hspace{5cm}           & $16.68_{-3.83}^{+6.88}$                            \\

sma ($R_{\odot}$)      &\hspace{5cm}           &  $1.20_{-0.13}^{+0.07}$                            \\

$M_1$ ($M_{\odot}$) &\hspace{5cm} & $0.29 \pm 0.02$              \\
$M_2$ ($M_{\odot}$)                  &\hspace{5cm}           & $0.53 \pm 0.05$                         \\
$f$(filling factor )                 &\hspace{5cm}           & $0.995\pm0.129$                   \\
\hline
\textbf{Light curve Modeling} &\hspace{5cm} &  \textbf{semi-detached model}   \\
$incl$($^{\circ}$)     &\hspace{5cm}           &$42.81_{-0.49}^{+0.63}$                    \\
$q(\frac{M_2}{M_1})$                 &\hspace{5cm}           &$1.87_{-0.14}^{+0.09}$               \\
$R_1$ ($R_{\odot}$) &\hspace{5cm} & $0.41 \pm 0.01$              \\

$relteff$           &\hspace{5cm}           & $0.78_{-0.11}^{+0.09}$                            \\
$colat$($^{\circ}$)     &\hspace{5cm}           & $107.98_{-7.68}^{+10.67}$                             \\
$long$($^{\circ}$)      &\hspace{5cm}           & $-126.56_{-3.34}^{+3.13}$                              \\
$radius$($^{\circ}$)            &\hspace{5cm}           & $16.87_{-4.15}^{+7.98}$                            \\

sma ($R_{\odot}$)      &\hspace{5cm}           &  $1.27_{-0.03}^{+0.02}$                            \\

$M_1$ ($M_{\odot}$) &\hspace{5cm} & $0.34 \pm 0.02$              \\
$M_2$ ($M_{\odot}$)                  &\hspace{5cm}           & $0.62 \pm 0.03$                         \\
\hline

\end{longtable}
\end{center}
}

\subsection{Flare events}
In Figure 1, it can be seen that there are flare events in each sector. We calculated the large flare energy in each sector. The bolometric energy of each flare is estimated using the following equation:
\begin{equation}
    E_{flare} = L_*\int_{}^{} \frac{F_f-F_q}{F_q} dt(erg)
\end{equation}
where $L_*$ is the stellar luminosity \citep{Shibayama+et+al+2013, Wu+et+al+2015, Chang+et+al+2017}, the $F_f$ and $F_q$ are the flaring and the quiescent flux. The Random Sample Consensus (RANSAC) \citep{Dao+et+al+2014} algorithm can eliminate some outliers and use the data from inliers for fitting. The process of calculating the large flare energy in Sector 23 is shown in Figure 11. The inlier points represent the quiescent flux, while the outlier points indicate the flare flux. A second-degree polynomial is used to fit the inlier points. The energy of large flare events in Sectors 23, 24, 50, and 51 is approximately $1.61 \times 10^{34}$ erg, $6.05 \times 10^{33}$ erg, $1.53 \times 10^{33}$ erg, and $5.25 \times 10^{35}$ erg, respectively.

\begin{figure}[!ht]
\begin{center}
\begin{minipage}{10cm}
	\includegraphics[width=10cm]{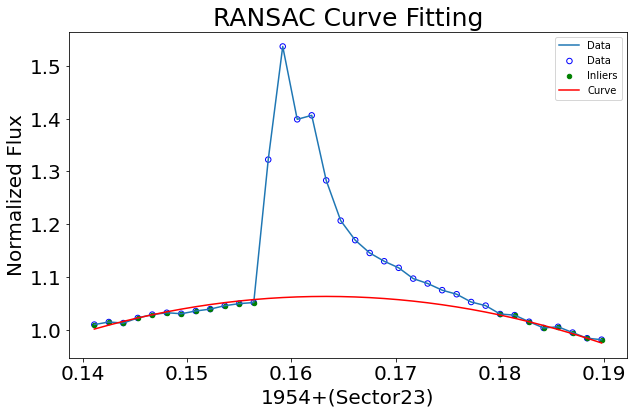}
\end{minipage}
\caption{The RANSAC algorithm is used to fit the quiescent flux. This algorithm effectively removes outliers and selects inliers for fitting.}\label{Fig10}
\end{center}
\end{figure}

\section{Conclusions and Discussions}
The International Variable Star Index \footnote{https://www.aavso.org/vsx/index.php} (VSX) classifies the star in question as a $\delta$ Scuti type. In this work, we demonstrate that the target is a binary system, with a red dwarf star and a compact object. Using spectra from LAMOST, the temperature of this red dwarf star is determined to be 3282K. The mass of red dwarf star is estimated to be $0.31\pm0.01$$M_{\odot}$ through SED fitting, and $0.29\pm0.02$$M_{\odot}$ using Phoebe with detached binary model, and $0.34\pm0.02$$M_{\odot}$ using Phoebe with semi-detached binary model base on the fitting of light curve and the radial velocity. These estimates indicate that the red dwarf star has both a low temperature and a small mass. In this target, specific manifestations of magnetic activity are observed in three aspects. Firstly, the light curve shows uneven heights, possibly indicating activity from cool spots. Secondly, the TESS observations show 4 distinct flare events in approximately 108 days of monitoring with the most energetic flare reaching up to $5.25 \times 10^{35}$ erg. Thirdly, strong H$\alpha$ emission lines are evident in the spectra obtained from LAMOST.

The H$\alpha$ emission line is doubled near orbital quadrature phase. A portion of the emission line is attributed to stellar activity of the M-dwarf star. We have attributed the additional narrow H$\alpha$ component that moves in antiphase with the M dwarf to the compact object, and our modeling supports the assumption that its velocity reflects the motion of the compact object. However, it is important to note that the physical origin of this emission is not fully understood. The narrowness of the line is intriguing, given the typically high orbital or infall speeds near a white dwarf. The mass ratio ($q$) is calculated to be $1.76 \pm 0.04$ based on the fitted radial velocities of two H$\alpha$ peaks. By fitting the SED to obtain the visible star mass ($M_1$) and combining it with the mass ratio ($q$) derived from radial velocity measurements, the mass of the compact star ($M_2$) is calculated to be $0.54 \pm 0.01$$M_{\odot}$. This value closely aligns with the compact star mass of $M_2 = 0.53 \pm 0.05$$M_{\odot}$ obtained using the Phoebe code with a detached binary model, as well as a slight difference in the compact star mass of $M_2 = 0.62 \pm 0.03$$M_{\odot}$ derived from the Phoebe code with a semi-detached binary model, based on the light curve and the radial velocity of the visible star. The minor discrepancy in $M_2$ could be linked to the accuracy of radial velocity measurements, which were based on a limited number of H$\alpha$ fitting points for the compact star. The similarity between the parameters derived from both the detached and semi-detached binary models can be attributed to the fact that, in the detached binary model, the Roche-lobe filling factor ($f$) of the M-dwarf star approaches 0.995, which is close to 1. Furthermore, the fitting results of the semi-detached binary model are also very good. These findings suggest that the M-dwarf star is likely undergoing continuous mass transfer. This compact star may potentially be a white dwarf, with its mass falling within the typical range of white dwarf masses.

\begin{acknowledgments}
We are very grateful for the data released by the TESS survey (https://archive.stsci.edu/missions-and-data/tess/) and LAMOST (http://www.lamost.org/public/). This work was partly supported by the Chinese Natural Science Foundation (No.12103088 and No.12433009),  National Key R\&D Program of China (No.2022YFF0711500 and No.2023YFA1608300), Yunnan Key Laboratory of Solar Physics and Space Science under the number 202205AG070009, Yunnan Provincial Foundation (No.202101AT070020), Yunnan Provincial Key Laboratory of Forensic Science  (No.YJXK005), Yunnan Basic Research Program (No.202201AU070116) and Yunnan Key Laboratory of Service Computing, Kunming, China. We acknowledge the science research grant from the China Manned Space Project with No.CMS-CSST-2021-A10, No.CMS-CSST-2021-A12 and No.CMS-CSST-2021-B10.

\end{acknowledgments}




\end{document}